# Electron-hole transport and photovoltaic effect in gated MoS$_2$ Schottky junctions

*Marcio Fontana[1,2], Tristan Deppe[1], Anthony K. Boyd[1], Mohamed Rinzan[1], Amy Y. Liu[1], Makarand Paranjape[1], and Paola Barbara[1]\*.*

[1] Department of Physics, Georgetown University, Washington, DC 20057, USA. [2] Department of Electrical Engineering, Federal University of Bahia, Salvador, 40210-630, Brazil.

\*e-mail: barbara@physics.georgetown.edu

**Semiconducting molybdenum disulfphide has emerged as an attractive material for novel nanoscale optoelectronic devices due to its reduced dimensionality and large direct bandgap. Since optoelectronic devices require electron-hole generation/recombination, it is important to be able to fabricate ambipolar transistors to investigate charge transport both in the conduction band and in the valence band. Although *n*-type transistor operation for single-layer and few-layer MoS$_2$ with gold source and drain contacts was recently demonstrated, transport in the valence band has been elusive for solid-state devices. Here we show that a multi-layer MoS$_2$ channel can be hole-doped by palladium contacts, yielding MoS$_2$ *p*-type transistors. When two different materials are used for the source and drain contacts, for example hole-doping Pd and electron-doping Au, the Schottky junctions formed at the MoS$_2$ contacts produce a clear photovoltaic effect.**

MoS$_2$ has a layered structure that can be mechanically exfoliated to produce thin flakes, similar to graphene[1, 2]. Bulk MoS$_2$ is a semiconductor with an indirect bandgap of about 1.2 eV[3]. When the thickness is reduced to a few layers, the indirect bandgap is tuned by quantum confinement and increases substantially, by 0.5 eV or more, until it eventually exceeds the energy spacing of the direct gap for single-layer thickness [4-6]. Recently fabricated MoS$_2$ transistors [7, 8] have shown promising characteristics for electronics applications. However, notwithstanding the potential advantages of a large direct bandgap for single-layer thickness, optoelectronic devices have not yet been explored, due to the difficulty in obtaining hole transport in MoS$_2$ transistors. While ambipolar MoS$_2$ transistors were recently reported, they either were gated with ionic liquids[9] or used PMMA as a substrate for MoS$_2$ [10]. In both cases, the mechanism leading to ambipolar transport in these samples is still unclear.

In this work, we demonstrate that all-solid-state devices that use standard SiO$_2$ as the MoS$_2$ substrate can show either *n*-type or *p*-type transistor behavior, depending on the choice of electrode material. When the source and drain contacts are both made of Au (with a thin Cr adhesion layer), our devices always show *n*-type behavior. In contrast, when source and drain contacts are both made of Pd (with a thin Nb adhesion layer), our $MoS_2$ transistors always show *p*-type behavior. Moreover, devices fabricated with two different contacts, namely Cr/Au and Nb/Pd each with thicknesses of 2nm/100nm, consistently show asymmetric ambipolar behavior and diode characteristics over a wide range of gate voltages. When illuminated, the devices with one Cr/Au and one Nb/Pd contact exhibit a sizable photovoltaic effect. All of these results can be consistently explained by considering the work functions for Pd, $MoS_2$, and Au, along with contributions to the band alignment due to interactions at the contact interfaces. In this picture, the photovoltaic effect arises from the built-in potential of the space charge accumulated at the source and drain contacts.

**Results**

Our samples are made of exfoliated multi-layer flakes of $MoS_2$ (about 50 nm thick) on a doped Si substrate capped with 300 nm $SiO_2$. Source and drain electrodes are patterned by e-beam lithography and deposited by sputtering. All measurements reported here were done at room temperature. Fig. 1 shows the transfer characteristics of two typical samples. In the case of Cr/Au source and drain electrodes (Fig. 1a), the device exhibits *n*-type behavior, similar to previous reports[7, 8]. Here we find that, even extending the measurements to a wide gate range of -100 V < $V_G$ < 100 V, no *p*-type conduction is measured. However, remarkably, in the case of Nb/Pd electrodes (Fig. 1b), the behavior is reversed and the sample shows only *p*-type conduction. This occurred consistently for all the ten devices measured.

We note that the Cr and Nb adhesion layers, for the Au and Pd contacts respectively, are extremely thin and do not form a continuous film. Therefore, the majority of the contact area between the flake and the source and drain electrodes will be formed by the $MoS_2$-Pd and $MoS_2$-Au interfaces. Nevertheless, we experimentally tested whether the Nb could be responsible for producing *p*-type conduction in our Nb/Pd samples. We did this by fabricating $MoS_2$ devices with both source and drain electrodes made with only Nb. The measured characteristics for all five devices indicated a clear and unambiguous *n*-type behavior, as shown in Fig 1c. This confirms that the *p*-type behavior in the Nb/Pd devices is solely due to the MoS$_2$-Pd interface. As a result, we will assume that the adhesion layers play an insignificant role in determining the device properties discussed in this work and we will refer to the Nb/Pd (Cr/Au) electrodes as Pd (Au).

To understand the contact dependence of the polarity, we first consider the work functions of the materials forming the contacts. Reported values of the work function for MoS$_2$, $\phi_{MoS2}$, range from 4.48 eV to 5.2 eV [11-14]. The literature values for polycrystalline Au and Pd work functions, $\phi_{Au}$ and $\phi_{Pd}$, are very similar, about 5.1 eV[15-17]. Based on the XPS core level shifts, the energy separation between the Fermi level and the valence band maximum at the Au-MoS$_2$ interface (i.e., the *p*-type Schottky barrier) is estimated to be 0.85 eV [14, 18, 19]. Our experimental results can be explained by considering an energy shift $\Delta\phi$ in the band alignment at the Pd-MoS$_2$ interface the lowers the Pd Fermi energy. This shift could be due to differences in the chemical interactions, charge redistribution, or Fermi level pinning[20] between the MoS$_2$-Pd and MoS$_2$-Au interfaces. For example, at the interface with Au, XPS data show that the Mo and S core levels shift by nearly the same amount, indicating a complete lack of chemical bonding between the Au and S layers[18]. At the interface with Pd, on the other hand, a small difference is measured between the Mo and S core level shifts, suggesting a stronger interaction at the interface[18]. Also, density functional theory investigations find that, at the Pd-MoS$_2$ interface, the Pd-S distance is slightly smaller than the sum of the covalent radii[21, 22], while, at the Au-MoS$_2$ interface, the Au-S distance is calculated to be somewhat larger[23]. While the ideal interfaces considered in those theoretical studies are not realized in our devices, the predicted trend of stronger interactions with Pd than with Au likely remains valid.

The band alignment for the Au-MoS$_2$ and Pd-MoS$_2$ interfaces is sketched in Fig. 2, where we assume that our MoS$_2$ flakes have electronic properties similar to that of the bulk, which has a bandgap of about 1.2 eV, and a work function on the larger end of the range of reported values. Since $E_{F, Pd} < E_{F, MoS2} < E_{F, Au}$, charge transfer occurs at the interfaces, causing doping of the MoS$_2$ channel and accumulation of space charge in the contact region, yielding Schottky barriers and either upward (hole doping) or downward (electron doping) bending of the conduction and valence band edges.

In the case of electron-doping Au contacts (Fig. 2a), when a positive gate voltage is applied, the bands shift downward, the curvature of the band profile is reversed and the Schottky barrier becomes thinner, yielding higher tunneling current through the conduction band. For negative gate voltages, the bands shift upward, the curvature of the band profile increases and large barriers at the contacts block the source-drain current through the valence band. An analogous picture can explain the *p*-type transfer characteristics of the Pd devices, assuming that the situation for Pd contacts is just the opposite (Fig 2b).

Moreover, since the work functions of both Nb and Cr are smaller than that of Au, this picture is consistent with the *n*-type behavior measured for devices with pure Nb contacts, and suggests that even if the adhesion metals play a non-negligible role in the device properties, their contributions should be *n*-type.

To further investigate the effect of different contact materials, we fabricated a device with two Pd electrodes and one Au electrode on the same MoS$_2$ flake, as illustrated in Figs. 3a,d. An optical image of the device is in Fig. 4a. This type of sample allows us to vary the combination of source and drain electrode material while using the same MoS$_2$ flake. When the two Pd contacts are used as source and drain electrodes, only *p*-type behavior is measured, as shown in Figs. 3e,f. This is consistent with two-electrode Pd-Pd devices fabricated with different flakes (e.g., see Fig. 1b). In contrast, when the Au contact is used as the drain electrode and either one of the Pd contacts is used as the source electrode, both *p*-type and *n*-type behavior can be measured, although the on-state current in the valence band is about one order of magnitude higher than the on-state current in the conduction band. Figs. 3c,d show the transfer characteristics for a bias configuration where the source electrode is the Pd contact furthest away from the Au contact. The curves are measured at equal and opposite values of source-drain voltage, $V_{SD}$, revealing hysteresis and a strongly asymmetric behavior as a function of source-drain bias.

Next, to explore the photoresponse of the device, we irradiated it with a 532 nm laser (photon energy about 2.3 eV, larger than the MoS$_2$ gap) with intensity of 1mW/mm$^2$. The laser spot size was 2 mm in diameter, which is much larger than the 2 μm spacing between source and drain electrodes; therefore in all measurements presented here, our devices were fully illuminated. We first explore the bias configuration in Fig. 3a, where one of the Pd electrodes and the Au electrode are used as source and drain contacts. For all values of gate voltage, when the sample is irradiated, the magnitude of the current through the device is *reduced* for positive source-drain bias and is *increased* for negative source-drain bias, as shown in Figs. 3b,c. Fig. 4b shows the current as a function of $V_{SD}$ (IV curve) at zero gate voltage, corresponding to the branch of the hysteresis with higher current, where the Fermi energy is shifted into the MoS$_2$ conduction band. The shape of the IV curve (black curve in Fig. 4b) is consistent with the band diagrams in Figs. 2c,d. For electron transport, the Schottky barrier at the Pd-MoS$_2$ interface yields Schottky-diode behavior. When a positive voltage is applied to the Pd electrode, it reduces the band bending, therefore lowering the barrier for electrons propagating from MoS$_2$ to Pd. This is the direct bias configuration and gives rise to a current increasing exponentially with the bias voltage. For source-drain bias of the opposite polarity, the current is limited to electrons thermally excited from Pd to MoS$_2$ over the Schottky barrier. When the device is irradiated by the laser, electrons from the Pd are photo-excited over the barrier and electron-hole pairs are created in the MoS$_2$ and separated by the built-in potential from the space charge at the contacts. Electrons accumulate on the Au side and holes accumulate on the Pd side, giving rise to an open circuit voltage, $V_{OC} = 0.1$ V, as shown by the red curve in Fig. 4b. Similar behavior is measured when the source voltage is applied to the Pd electrode closer to the Au electrode and with other flakes on different chips having only two contacts, one Pd and one Au (Figs. 5a,b).

By contrast, the photoresponse from devices with Pd-Pd contacts is qualitatively very different, as shown in Figs. 3e,f for the Pd-Pd bias configuration of the three-contact device, and in Figs. 5c,d for a different flake with only two Pd-Pd contacts. The curves are more symmetric with respect to $V_{SD}$, and regardless of the polarity of $V_{SD}$, the magnitude of the current always *increases* when the laser is on. In this case, the built-in potentials at the two ends of the channel accelerate charges in opposite directions (Fig. 2b); therefore there is no net charge separation across the open-circuit device and $V_{OC} = 0$. However, under laser irradiation, the increased number of charge carriers due to the creation of electron-hole pairs increases the magnitude of the current through the device for both polarities of $V_{SD}$. Similar photoresponse has been reported for single-layer $MoS_2$ transistors with Au contacts[24].

**Discussion**

We have demonstrated polarity control in multilayer $MoS_2$ solid-state transistors. In particular, we find that the choice of material for the source and drain electrodes is key to controlling whether transport occurs through the conduction or valence band. When two different materials are used for the source and drain electrodes, one hole-doping and one electron-doping, both *n*-type and *p*-type behavior can be measured. In addition, our gated $MoS_2$ Schottky diodes show a photovoltaic effect. We estimate that the maximum electrical power that can be extracted from the device is about 2.5% of the laser power incident on the $MoS_2$ region between the electrodes. This percentage is a lower bound for the conversion efficiency, because only the depletion region in the $MoS_2$ channel is photoactive. Nevertheless, this room-temperature efficiency is better than the highest value previously reported in $MoS_2$, about 1% at 120K for bulk samples[25]. This is a significant effect, given that our devices were made of multilayer $MoS_2$, which has an indirect bandgap. Much higher conversion efficiency is expected for similar diodes made of single-layer $MoS_2$, which has a direct bandgap. These results show that metal contact engineering is a very promising avenue for building $MoS_2$-based solid-state optical devices, with fabrication techniques that can be scaled into dense arrays with potential applications in flexible optoelectronics. Future work will involve the study of these devices with a laser spot size smaller than the spacing between the source and drain electrodes, in order to investigate the photocurrent as a function of laser light position.

**Methods**

*Fabrication and characterization of MoS$_2$ devices:* $MoS_2$ flakes were exfoliated[1] from bulk $MoS_2$ by SPI Supplies using scotch tape. Prior to the mechanical exfoliation procedure, the silicon wafer (boron doped silicon with 300 nm $SiO_2$ grown by thermal dry oxidation) was cut into chips measuring 10 mm x 10 mm, which were ultrasonically cleaned in trichloroethylene ($C_2HCl_3$), acetone (($CH_3)_2CO$) and isopropyl alcohol ($C_3H_8O$), rinsed in deionized water and dried in nitrogen. Alignment marks were patterned, using electron beam lithography (EBL) and lift-off. After imaging $MoS_2$ flakes with a field emission scanning electron microscope (FESEM) by Zeiss Microscope, source/drain electrodes are patterned by EBL. The width of the source and drain metal electrodes are 2 μm with a 2 μm separation. The metals (Cr/Au and Nb/Pd with thicknesses of 20 Å/1000 Å) are deposited by sputtering and followed by lift-off in acetone.

The thickness of the $MoS_2$ flakes was measured with an atomic force microscope, JEOL scanning probe microscope JSPM-4200, using non-contact mode.

The electrical characterization of the devices was done with a room temperature probe station, consisting of three micromanipulators to contact the sample and a combined pA meter – dual DC voltage source (HP4140B) controlled by a LabVIEW program.

**Acknowledgments**

This work was supported by the NSF (DMR-1008242, DMR-1006605). M. Fontana acknowledges support from the Brazilian National Council for Scientific and Technological Development (CNPq). The authors thank E. Van Keuren for assistance with the optical set-up.

**Author Contributions**

Device fabrication and transport measurements were performed by M.F.

T.D. and A.K.B. assisted with the fabrication and M.R. with the electrical characterization.

M.F., P.B., A.Y.L. and M.P. interpreted the data.

All authors discussed the results and contributed to the manuscript preparation.

**Additional Information**

The authors declare no competing financial interests.

**Figure 1. Transfer characteristics of MoS$_2$ transistors.** (a) MoS$_2$ transistor with Au source and drain contacts. Only *n*-type behavior is observed. An optical image of the device is shown in the inset. (b) MoS$_2$ transistor with Pd source and drain contacts. Only *p*-type behavior is observed. (c) MoS$_2$ transistor with Nb source and drain contacts. Only *n*-type behavior is observed.

**Figure 2. Schematic band alignment for different source and drain electrode materials.** (a) Two Au electrodes. (b) Two Pd electrodes. (c) One Au and one Pd electrode. In (a), (b), and (c), the left panel shows the different band alignment for Pd and Au discussed in the text and the right panel illustrates the corresponding band bending after contact. The shift in bands upon application of a gate voltage is also shown. (d) Effect of source-drain bias on device with one Au and one Pd electrode.

**Figure 3. MoS$_2$ flake with three-contacts.** (a-c) Transfer characteristics and photoresponse for positive (b) and negative (c) source-drain bias, corresponding to the Pd-Au bias configuration shown in (a). (d-f) Transfer characteristics and photoresponse for positive (e) and negative (f) source-drain bias, corresponding to the Pd-Pd bias configuration shown in (d).

**Figure 4. Photovoltaic effect with Pd-Au bias configuration.** (a) Optical image of the device. The spacing between the electrodes is 2 μm. (b) Current vs. source-drain voltage at $V_G = 0$ showing strong asymmetry and photoresponse with diode-like behavior for the Pd-Au bias configuration indicated in (a).

**Figure 5. Photoresponse from two-contact devices.** (a) Schematic of device with one Pd and one Au contact. (b) Current vs. source-drain voltage for device with Pd-Au contacts, with and without laser illumination. (c) Schematic of device with two Pd contacts. (d) Current vs. source-drain voltage for device with Pd-Pd contacts, with and without laser illumination.

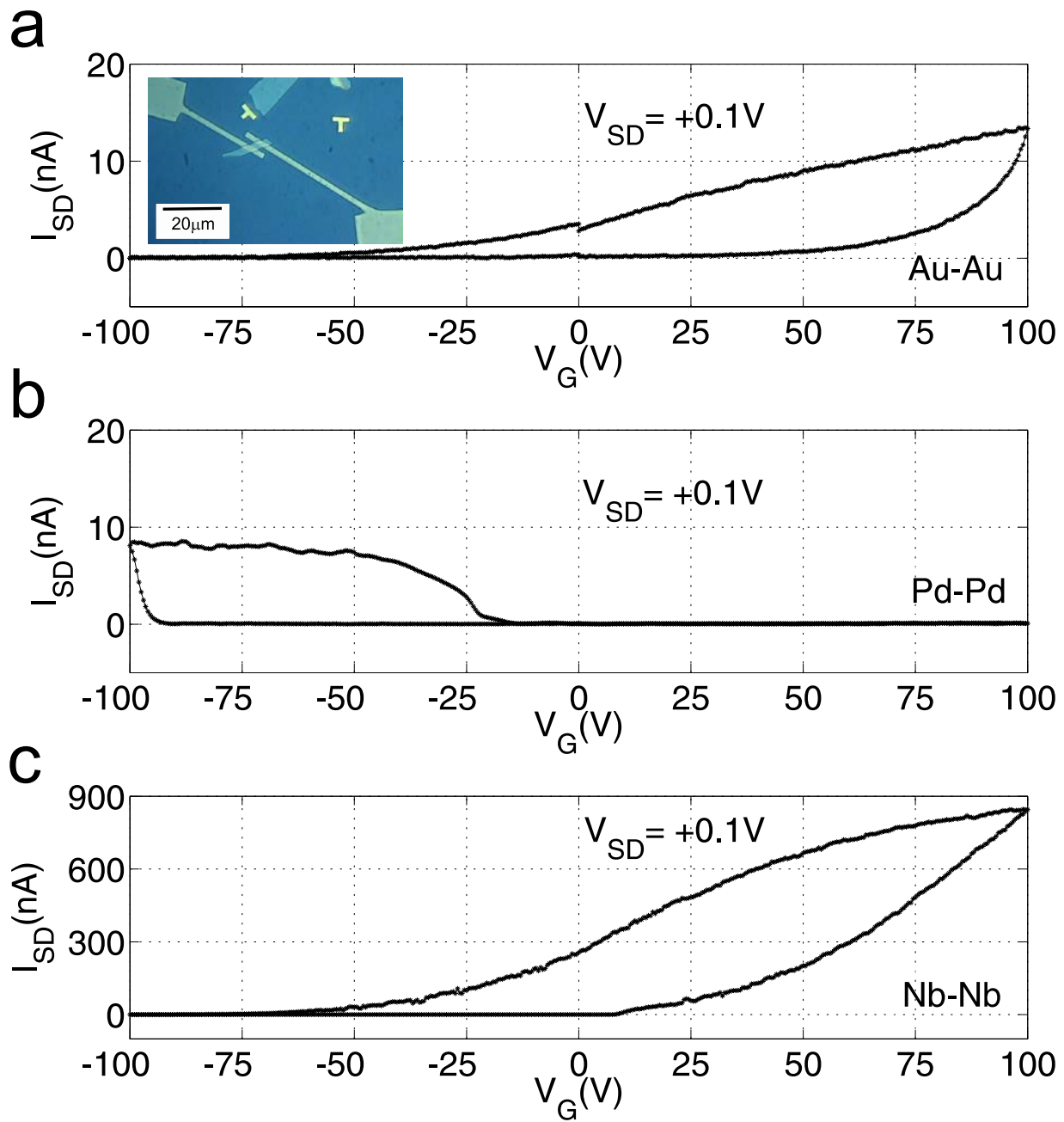

**Figure 1. Transfer characteristics of MoS$_2$ transistors.** (a) MoS$_2$ transistor with Au source and drain contacts. Only *n*-type behavior is observed. An optical image of the device is shown in the inset. (b) MoS$_2$ transistor with Pd source and drain contacts. Only *p*-type behavior is observed. (c) MoS$_2$ transistor with Nb source and drain contacts. Only *n*-type behavior is observed.

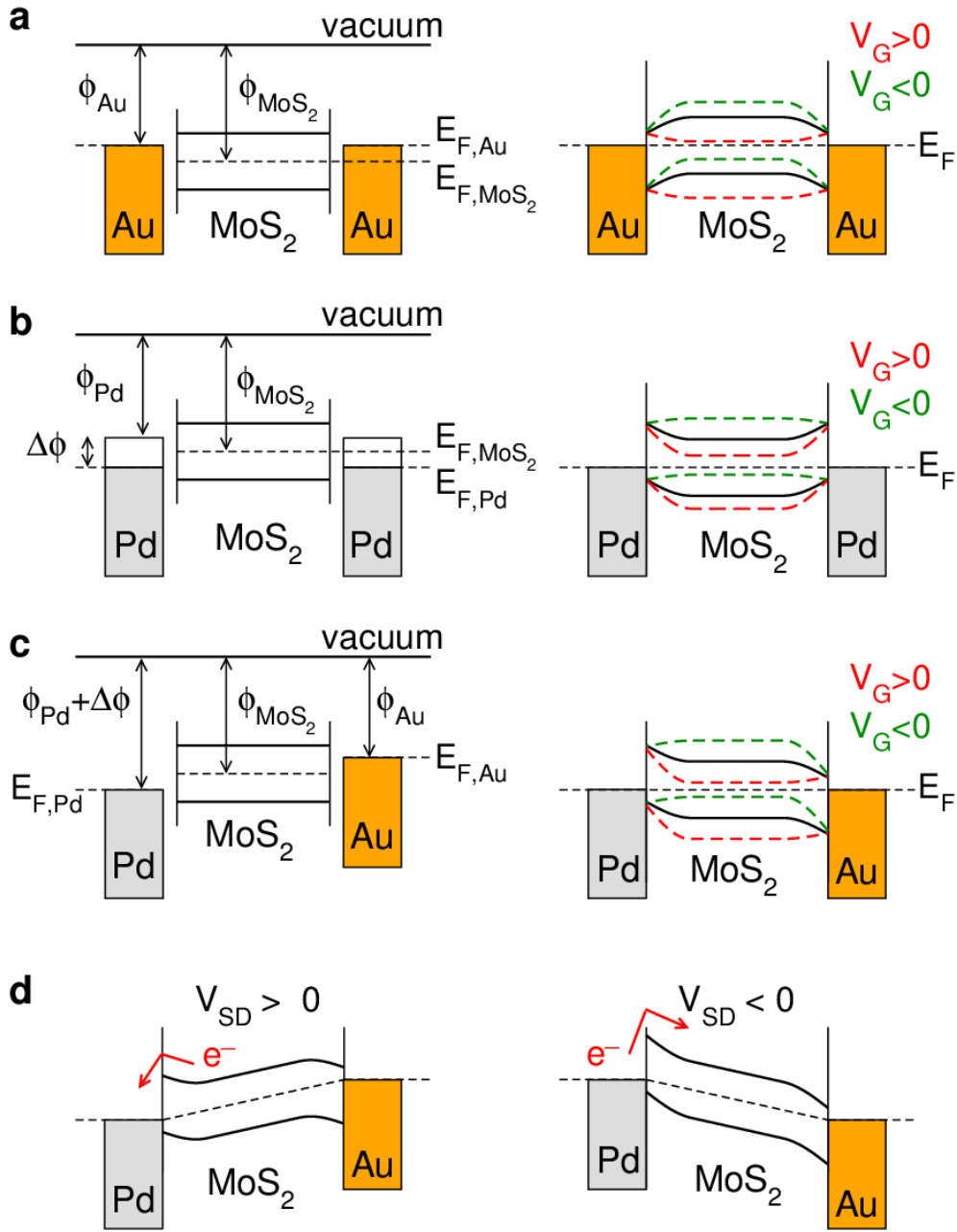

**Figure 2. Schematic band alignment for different source and drain electrode materials.** (a) Two Au electrodes. (b) Two Pd electrodes. (c) One Au and one Pd electrode. In (a), (b), and (c), the left panel shows the different band alignment for Pd and Au discussed in the text and the right panel illustrates the corresponding band bending after contact. The shift in bands upon application of a gate voltage is also shown. (d) Effect of source-drain bias on device with one Au and one Pd electrode.

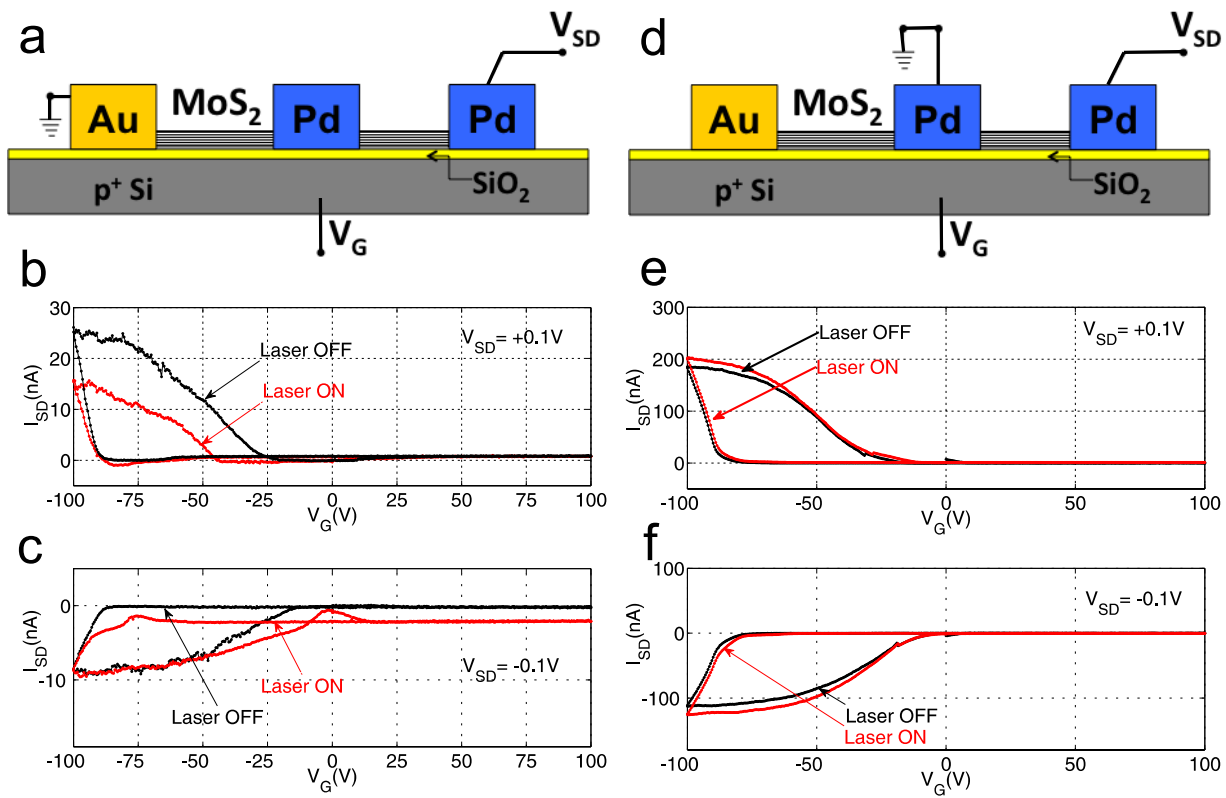

**Figure 3. MoS$_2$ flake with three-contacts.** (a-c) Transfer characteristics and photoresponse for positive (b) and negative (c) source-drain bias, corresponding to the Pd-Au bias configuration shown in (a). (d-f) Transfer characteristics and photoresponse for positive (e) and negative (f) source-drain bias, corresponding to the Pd-Pd bias configuration shown in (d).

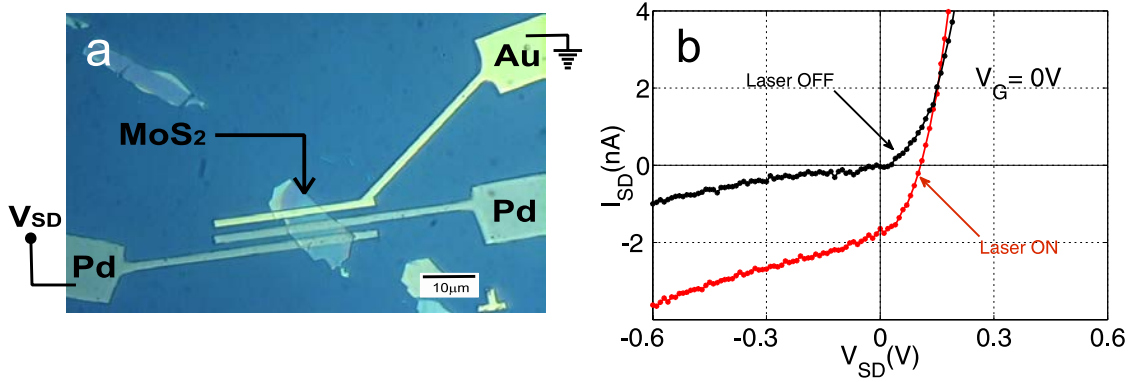

**Figure 4. Photovoltaic effect with Pd-Au bias configuration.** (a) Optical image of the device. The spacing between the electrodes is 2 μm. (b) Current vs. source-drain voltage at $V_G = 0$ showing strong asymmetry and photoresponse with diode-like behavior for the Pd-Au bias configuration indicated in (a).

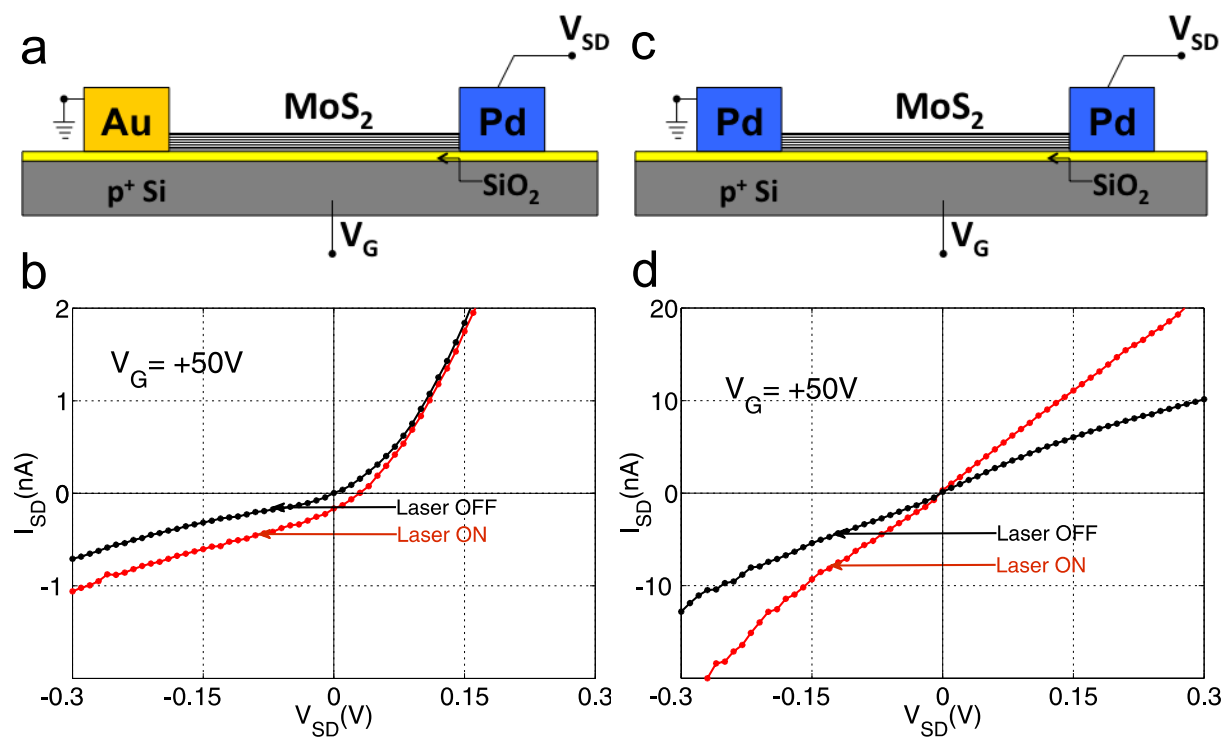

**Figure 5. Photoresponse from two-contact devices on different MoS$_2$ flakes.** (a) Schematic of device with one Pd and one Au contact. (b) Current vs. source-drain voltage for device with Pd-Au contacts, with and without laser illumination. (c) Schematic of device with two Pd contacts. (d) Current vs. source-drain voltage for device with Pd-Pd contacts, with and without laser illumination.